\documentclass[printer]{tCPH2ep}
\usepackage{graphicx}
\usepackage{wrapfig}
\usepackage[numbers]{natbib}
\usepackage{hyperref}

\usepackage{color}
\definecolor{IITred}{rgb}{0.5,0.05,0.05}
\newcommand{\query}[1]{\textcolor{IITred}{\textbf{#1}}}
\renewcommand{\query}[1]{#1}
\newcommand{\cf}{{\em cf.\ }}

\newcommand{\gev}{\hbox{ GeV}}

\newcommand{\mev}{\hbox{ MeV}}

\newcommand{\tev}{\hbox{ TeV}}

\newcommand{\s}{\hbox{ s}}

\newcommand{\cm}{\hbox{ cm}}

\newcommand{\pb}{\hbox{ pb}}
\newcommand{\fb}{\hbox{ fb}}

\newcommand{\m}{\hbox{ m}}

\newcommand{\alphas}{\ensuremath{\alpha_{\mathrm{s}}}}

\newcommand{\ewgg}{\ensuremath{\mathrm{SU(2)_L }\otimes \mathrm{U(1)}_Y}}
\newcommand{\wigg}{\ensuremath{\mathrm{SU(2)_L }}}
\newcommand{\ygg}{\ensuremath{\mathrm{U(1)}_Y}}

\begin{document}

\markboth{Chris Quigg}{After Higgs}

\title{Particle Physics after the Higgs-Boson Discovery: \\ {Opportunities for the Large Hadron Collider}}

\author{Chris Quigg$^{\ast}$\thanks{$^\ast$Email: quigg@fnal.gov \hfill \textsf{FERMILAB--PUB--15/290--T}}\\[6pt]
  \em{Theoretical Physics Department, Fermi National Accelerator Laboratory\\ P.O. Box 500, Batavia, Illinois 60510 USA}\\[4pt] Institut de Physique Th\'eorique Philippe Meyer\\ Laboratoire de Physique Th\'eorique de l'\'Ecole Normale Sup\'erieure\\ 24 rue Lhomond, 75231 Paris Cedex 05, France\\[6pt]
\received{Revised August 14, 2015}}
\maketitle

\begin{abstract}
The first run of the Large Hadron Collider at CERN brought the discovery of the Higgs boson, an apparently elementary scalar particle with a mass of $125\gev$, the avatar of the mechanism that hides the electroweak symmetry. A new round of experimentation is beginning, with the energy of the proton--proton colliding beams raised to $6.5\tev$ per beam, from $4\tev$ at the end of the first run. This article summarizes what we have learned about the Higgs boson, and calls attention to some issues that will be among our central concerns in the near future. 

\end{abstract}

\section{Introduction \label{sec:Intro}}
In a previous contribution~\cite{Quigg:2007fj}, written before experiments began at the Large Hadron Collider, I laid out the essential elements of the electroweak theory and prospects for the first wave of experimental studies at energies around $1\tev$ for collisions among quarks and gluons.\footnote{For a more expansive prospectus, still relevant today, see Ref.~\cite{{Quigg:2009vq}}.} In that regime, we could confidently expect to learn about the mechanism that hides the gauge symmetry on which the electroweak theory is founded. Although the LHC operated at less than 60\% of its $7 \oplus 7\tev$ design energy, experiments in 2011--2012 took the first giant step toward understanding electroweak symmetry breaking, discovering the Higgs boson and beginning to catalogue its properties.

Now that protons are colliding in the LHC at a record energy of $6.5\tev$ per beam, launching a new wave of experimentation, it is timely to summarize what we have learned about the Higgs boson, what remains to be learned, and what questions are raised by the observed value of the Higgs-boson mass. Looking beyond the symmetry-breaking sector, the discovery of the Higgs boson, together with other  observations, sets a new context for a range of concerns in particle physics. Accordingly, the second aim of this paper will be to identify several different areas in which we can expect new and definitive information from the LHC experiments in the coming few years. My discussion focuses on ``hard-scattering'' interactions at the energy frontier that are the domain of the ATLAS and CMS experiments. A comprehensive survey of particle physics in 2015 would also address the unreasonable effectiveness of the standard model, concentrating on flavor physics, properties and interactions of neutrinos, and the study of hadronic matter under extreme conditions.

\section{The Higgs Boson \label{sec:higgs}}
The Higgs boson has been the object of one of the greatest campaigns in the history of particle physics and a pop-culture icon. On 4 July 2012, the ATLAS~\cite{Aad:2012tfa}
and CMS~\cite{Chatrchyan:2012ufa} Collaborations announced the discovery of a Higgs-boson candidate with mass near $125\gev$.\footnote{An accessible summary of the experiments is given in Ref.~\cite{DellaNegra:2012mga}. Various aspects of the theoretical context and of the search and discovery are chronicled in a number of recent books~\cite{Massive,Infinity,Sean,Lisa,Beyond,Butterw}.} ATLAS and CMS searched for decays of a Higgs boson into $b$-quark--antiquark ($b\bar{b}$) pairs, tau-lepton ($\tau^+\tau^-$) pairs, and pairs of electroweak gauge bosons: $W^+W^-$ or $Z^0Z^0$, where one or both of the weak bosons may be virtual, or two photons ($\gamma\gamma$). By now, signals for all these modes have been identified. The primary discovery evidence came from peaks observed in the $\gamma\gamma$ and four-lepton ($Z^0Z^0 \to \ell^+\ell^-\ell^{\prime+}\ell^{\prime-}$, where $\ell$ and $\ell^\prime$ can denote an electron ($e$) or muon ($\mu$))  invariant-mass distributions.
Subsequent studies confirm that the new particle's properties closely match those of the standard-model Higgs boson. Before we review those properties in detail, let us take a moment to recall what a Higgs boson is, and what role it plays in the standard model of particle physics.
\subsection{The Higgs boson in the Electroweak Theory \label{subsec:HiggsEW}}
The basic constituents of the standard model are six flavors each of quarks and leptons, spin-$\frac{1}{2}$ particles that are structureless at the current limits of resolution, about $10^{-19}\m$. As Dirac fermions, the quarks and charged leptons have distinct left-handed and right-handed chiral components. The character of the neutrinos has not yet been fully clarified, but it is established that the neutrinos that participate in the observed weak interactions are purely left-handed. For the purposes of this article, we can set aside the phenomenon of neutrino oscillations.\footnote{For an excellent brief introduction, including reprints of some classic papers, see Ref.~\cite{CahnGold}, Chapter 19.}

As C.~N.~Yang has observed~\cite{YangSym}, the twentieth century saw a profound change in the place of symmetry in the physicist's conception of nature. A passive role, in which symmetry was seen as a property of interactions, has been supplanted by an active role within the gauge-theory paradigm: \emph{symmetry dictates interaction.}

The electroweak theory~\cite{Glashow:1961tr,Weinberg:1967tq,Salam:1968rm} is built on a left-handed weak-isospin symmetry, \wigg, reflecting the fact that charge-changing weak interactions involve only the left-handed quarks and leptons, and a weak-hypercharge phase symmetry, \ygg.\footnote{This choice for the gauge group was arrived at by trial and error, not by derivation from a guiding principle.} In a gauge theory, the interactions are mediated by spin-1 gauge bosons. The \query{combined} \ewgg\ symmetry implies four massless gauge bosons: a weak isotriplet arising from \wigg\ and a hyperphoton arising from \ygg. Although the theory captures symmetries abstracted from experiment, it does not describe the real world: Weak bosons must be massive---because the range of weak interactions is restricted to about $10^{-17}\m$ and the only observed long-range force is electromagnetism, mediated by the photon. \query{In Dirac's theory, which underlies modern gauge theories, a spin-$\frac{1}{2}$ fermion such as the electron have distinct left-handed and right-handed components. A term $m\bar{e}e$ in the Lagrangian leads, through the Euler--Lagrange equations, to the familiar mass term in the Dirac equation. Expanding $m\bar{e}e = m\bar{e}[\frac{1}{2}(1 - \gamma_5) + \frac{1}{2}(1 + \gamma_5)]e = m(\bar{e}_{\mathrm{R}}e_{\mathrm{L}} + \bar{e}_{\mathrm{L}}e_{\mathrm{R}})$, we see that a mass term joins the left-handed and right-handed components. Because the left-handed and right-handed fermion fields transform differently, such a mass term would not respect the gauge symmetry.} 

A great insight of twentieth-century science is that symmetries expressed in the laws of nature need not be manifest in the outcomes of those laws. Condensed-matter physics is rich in examples. Of special relevance to the challenge of massless gauge fields is the Meissner effect, the exclusion of magnetic flux from a superconductor. The photon acquires mass within the superconducting medium, limiting the penetration depth of the magnetic field; the gauge invariance of electromagnetism is hidden~\cite{Anderson:1958pb}. The application of these analogues to quantum field theories based on gauge symmetry was carried out in the mid-1960s~\cite{Englert:1964et,Higgs:1964ia,Higgs:1964pj,Guralnik:1964eu,Higgs:1966ev,Kibble:1967sv}. It is for this work that Fran\c{c}ois Englert~\cite{Englert:2014zpa} and Peter Higgs~\cite{Higgs:2014aqa} received the 2013 Nobel Prize for Physics. 

Exploiting the discovery of how spontaneous symmetry breaking operates in gauge theories, Weinberg~\cite{Weinberg:1967tq} and Salam~\cite{Salam:1968rm} showed how to give masses to the gauge bosons and constituent fermions, within the \ewgg\ framework.\footnote{For an essay on the diverse threads that came together in the electroweak theory, see Ref.~\cite{Quigg:2015cfa}. Recent textbook treatments include Refs.~\cite{Langacker:2010,MattS,CQGT2,DynSM}.}   The circumstances that hide the gauge symmetry, by producing a degenerate set of vacuum states that do not respect the symmetry, are the work of a quartet of scalar fields. The Weinberg--Salam construction implied a new class of \query{charge-preserving} weak interactions, the weak-neutral-current interactions mediated by the $Z^0$ gauge boson, predicted the masses of $Z^0$ and the mediator of the charged-current weak interaction, $W^\pm$, and demonstrated that the masses of the quarks and charged leptons could arise from interactions with the scalar fields introduced to hide the gauge symmetry. Three of the four scalar degrees of freedom become the longitudinal components of $W^\pm$ and $Z^0$. The remaining scalar degree of freedom emerges as the Higgs boson of the electroweak theory. The electroweak theory does not predict a specific value for the Higgs-boson mass.

Neutral currents (observed first in $\bar{\nu}_\mu e \to \bar{\nu}_\mu e$ scattering) were discovered in 1973,\footnote{For a first-person account of the discovery by the Gargamelle experiment and others, see Ref.~\cite{Haidt:2004ne}.} the $W^\pm$ and $Z^0$ a decade later.\footnote{See Carlo Rubbia's Nobel Lecture, Ref.~\cite{Rubbia:1985pv}, for an account of the search by the UA1 and UA2 Collaborations.} By the turn of the twenty-first century, the electroweak theory, augmented to include six quark flavors and six lepton flavors, had been validated by more than a dozen measurements at a precision of a few parts per thousand~\cite{Z-Pole,LEP-2}. The Higgs boson remained missing, although its virtual effects had been detected in quantum corrections. The role of the Higgs field in giving masses to fermions was untested.

\subsection{What Experiments Have Revealed \label{subsec:revealed}}
After the discovery of neutral currents and the demonstration that spontaneously broken gauge theories can be renormalized~\cite{'tHooft:2000xn,Veltman:2000xp}, the Higgs boson became an object of desire~\cite{Ellis:1975ap} for particle-physics experiments.  A partial-wave unitarity analysis of $W^+W^-$ scattering and related processes at high energies \query{(\cf \S4 of Ref.~\cite{Quigg:2007fj})} showed that either a standard-model Higgs boson would be found with mass $M_H \lesssim 1\tev$ or other new phenomena would occur on the TeV scale~\cite{Lee:1977eg}. The most extensive searches were carried out at the Large Electron--Positron Collider at CERN, in studies of the reaction $e^+ e^- \to HZ^0$~\cite{Kado:2002er}, which established a lower-bound on the mass of a standard-model Higgs boson, $M_H > 114.4\gev$ at 95\% confidence level~\cite{Barate:2003sz}. The detailed data analyses that validated the electroweak theory, exemplified by Refs.~\cite{Z-Pole,LEP-2}, indicated that quantum corrections from the Higgs boson were needed, within the standard-model framework, and that the standard-model Higgs boson should be relatively light.\footnote{To trace the evolution from 1996 to 2012 of the LEP Electroweak Working Group's ``blueband plots,'' showing the goodness of fit as a function of the Higgs-boson mass, consult \url{http://lepewwg.web.cern.ch/LEPEWWG/plots/}.} The Tevatron Collider experiments at Fermilab, in which protons and antiprotons collided at c.m.\ energy $\sqrt{s} = 1.96\tev$, conducted searches in a large number of channels. The CDF and D0 experiments were particularly sensitive to decays into pairs of weak gauge bosons. By July 2012, their joint analysis excluded the range $145\gev \le M_H \le 180\gev$ at 95\% CL~\cite{Group:2012zca}.\footnote{A review of Higgs-boson searches at the Tevatron is given in Ref.~\cite{Junk:2014yea}.}

While the discovery of a standard-model Higgs boson was widely anticipated, and the evidence that accumulated in advance of the discovery encouraged that belief, an elementary scalar was not a foregone conclusion. It remained at least as a logical possibility  that the electroweak symmetry might be  broken dynamically or be related through extra spacetime dimensions to gravity, to cite two leading options. The alternatives are now strongly disfavored as the dominant element of electroweak symmetry breaking.

\begin{figure}
\centerline{\includegraphics[width=0.45\textwidth]{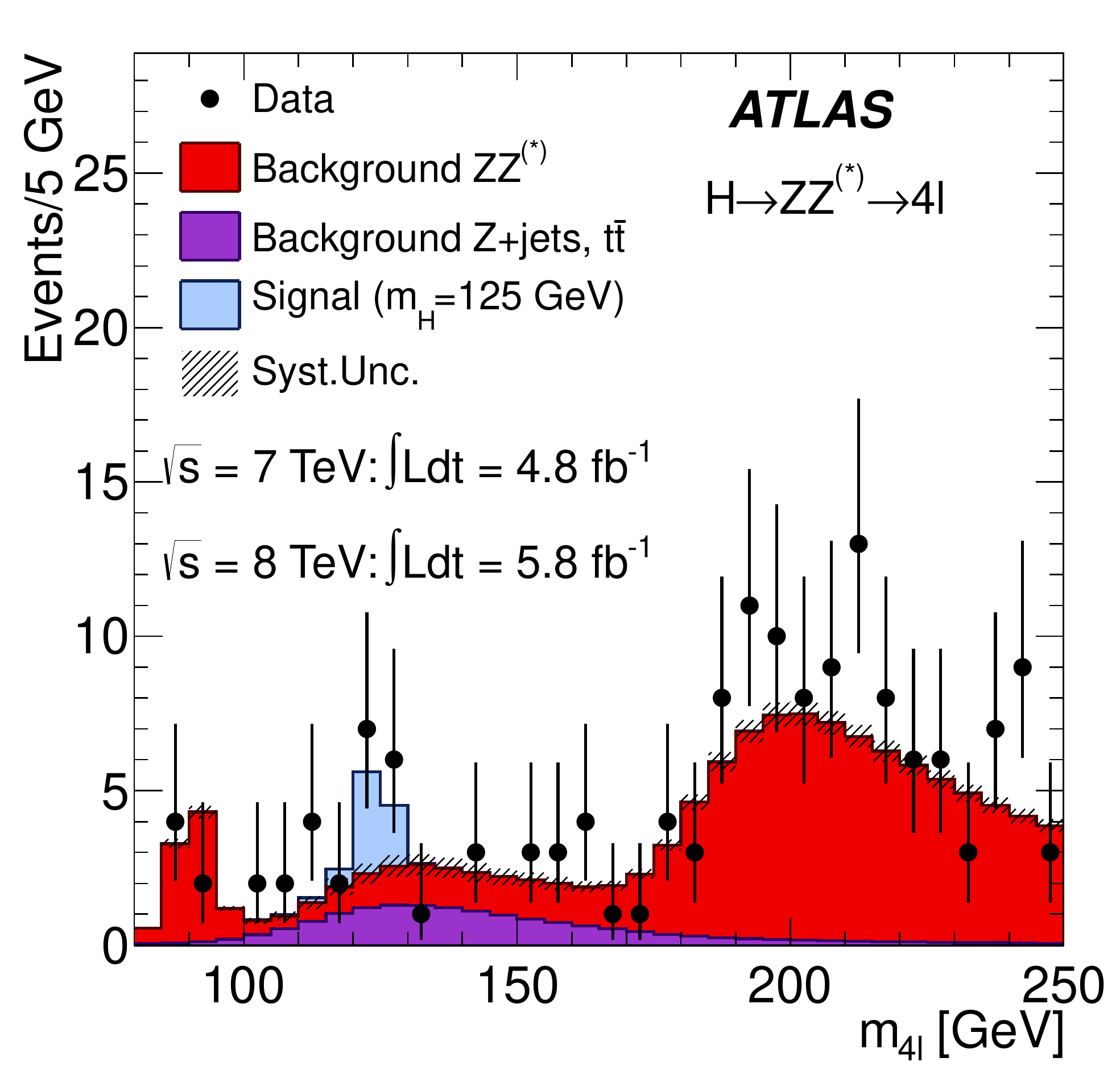}\quad\includegraphics[width=0.4571\textwidth]{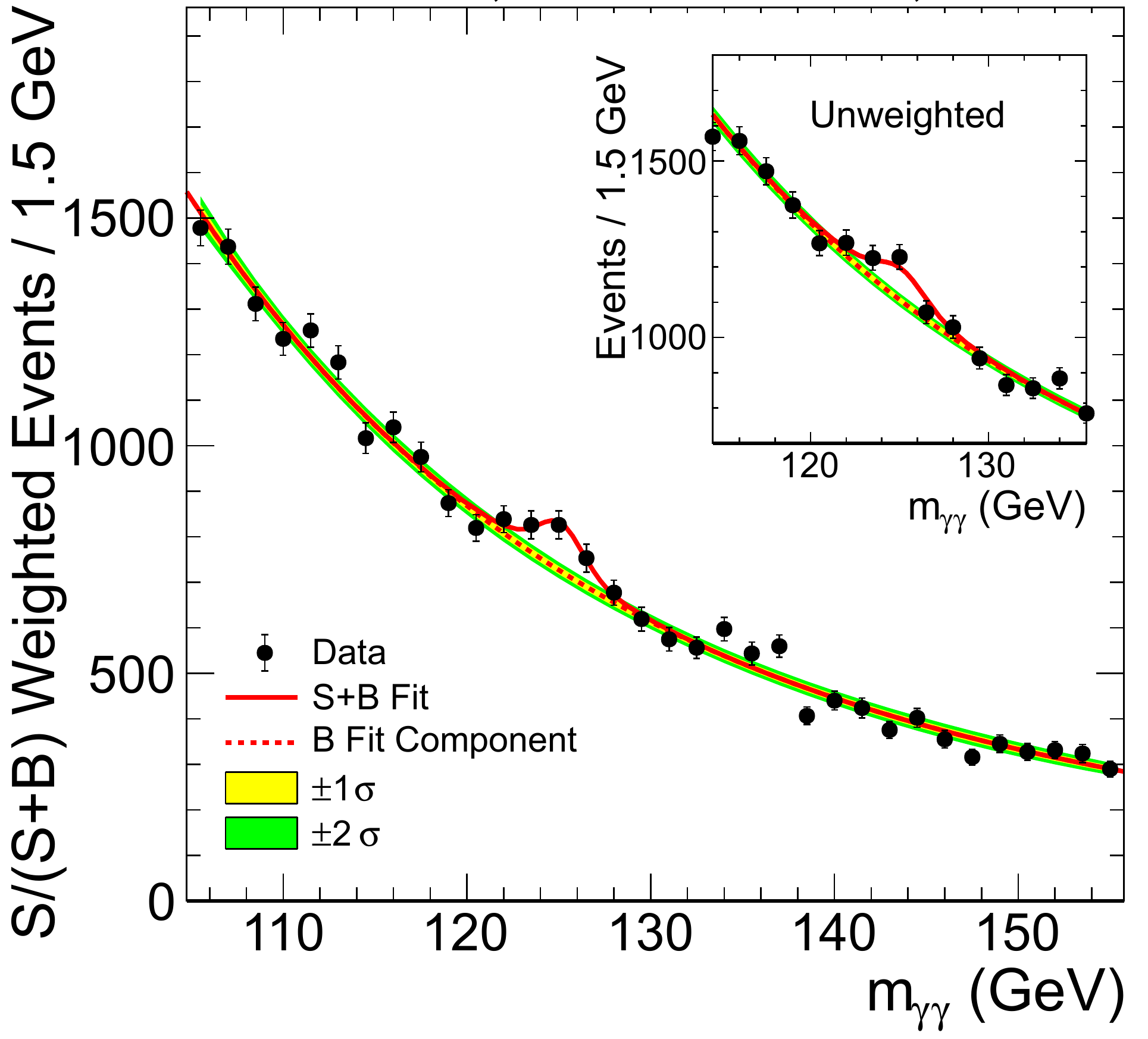}}
\caption{Two examples of the discovery evidence from combined data at $\sqrt{s}= 7\hbox{ and }8\tev$. Left panel: The observed four-lepton invariant mass, compared to expected backgrounds,  in the 80 to $250\gev$ mass range in the ATLAS Experiment~\cite{Aad:2012tfa}. The signal expectation for a standard-model Higgs boson with $M_H = 125\gev$ is also shown. Right panel: The diphoton invariant mass distribution observed by the CMS Experiment~\cite{Chatrchyan:2012ufa}, with each event weighted by the Signal/(Signal + Background) value of its category. The lines represent the fitted background and signal, and the colored bands represent  $\pm (1,2)\sigma$ uncertainties in the background estimate. The inset shows the central part of the unweighted invariant mass distribution.  \label{fig:Hdisc}}
\end{figure}

ATLAS~\cite{ATLASdet} and CMS~\cite{CMSdet} are large, broad-acceptance, general-purpose detectors located in multistory caverns about 100 meters below the surface, in the Large Hadron Collider tunnel. In the discovery papers, ATLAS~\cite{Aad:2012tfa} and CMS~\cite{Chatrchyan:2012ufa} analyzed approximately $5\fb^{-1}$ integrated luminosity\query{\footnote{\query{Luminosity, expressed in $\cm^{-2}\s^{-1}$, is a convenient measure of collision rate for colliders. When multiplied by the relevant cross section, it gives the event rate per second. Integrated luminosity, expressed as an inverse area, is a time integral of luminosity. When multiplied by a cross section, it yields the number of events.}}} collected at $\sqrt{s} = 7\tev$ in 2011 and a similar amount at $\sqrt{s} = 8\tev$ in 2012. The clearest signals, narrow invariant-mass peaks in the $\gamma\gamma$ and four-lepton ($Z^0Z^0 \to \ell^+\ell^-\ell^{\prime+}\ell^{\prime-}$) channels, signaled the production and decay of a new particle with mass around $125\gev$. Examples of the discovery signals are given in Figure~\ref{fig:Hdisc}.
A joint analysis, based on all the Run~1 data for these two ``high-resolution'' channels, yields $M_H = 125.09 \pm 0.24\gev$~\cite{Aad:2015zhl}. At the time of discovery, the CDF and D0 Collaborations reported a $3.1\sigma$ excess of $b\bar{b}$ pairs, consistent with the production of a 125-GeV Higgs boson, in association with a $W^\pm$ or $Z^0$~\cite{Aaltonen:2012qt}. ATLAS and CMS have also observed the new particle in the $W^+W^-$ mode~\cite{Chatrchyan:2013iaa,ATLAS:2014aga}, and have given evidence for the $b\bar{b}$~\cite{Khachatryan:2015bnx} and $\tau^+\tau^-$~\cite{Aad:2015vsa,Chatrchyan:2014nva} modes. 

\begin{figure}
\centerline{\includegraphics[height=0.3\textheight]{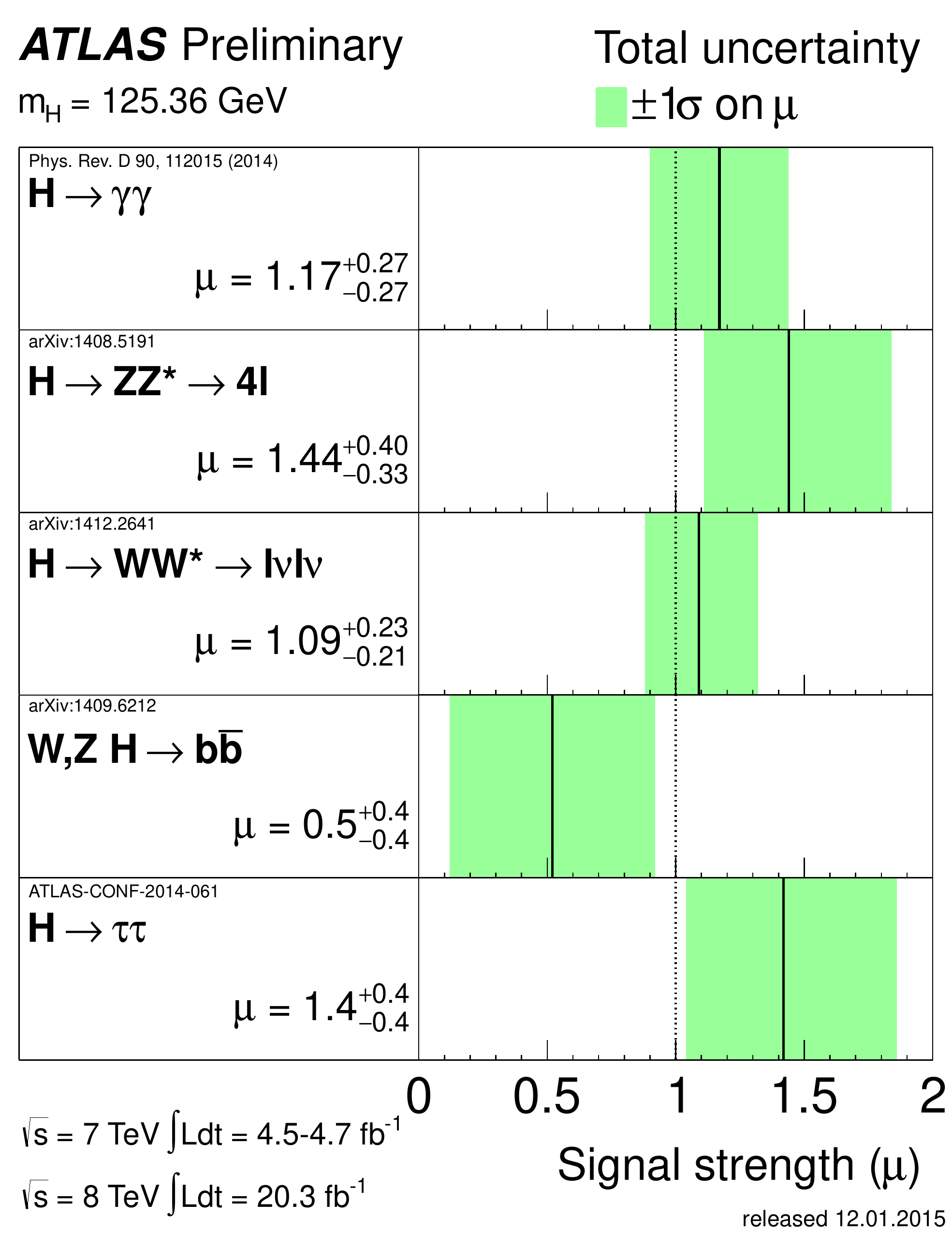}\includegraphics[height=0.3\textheight]{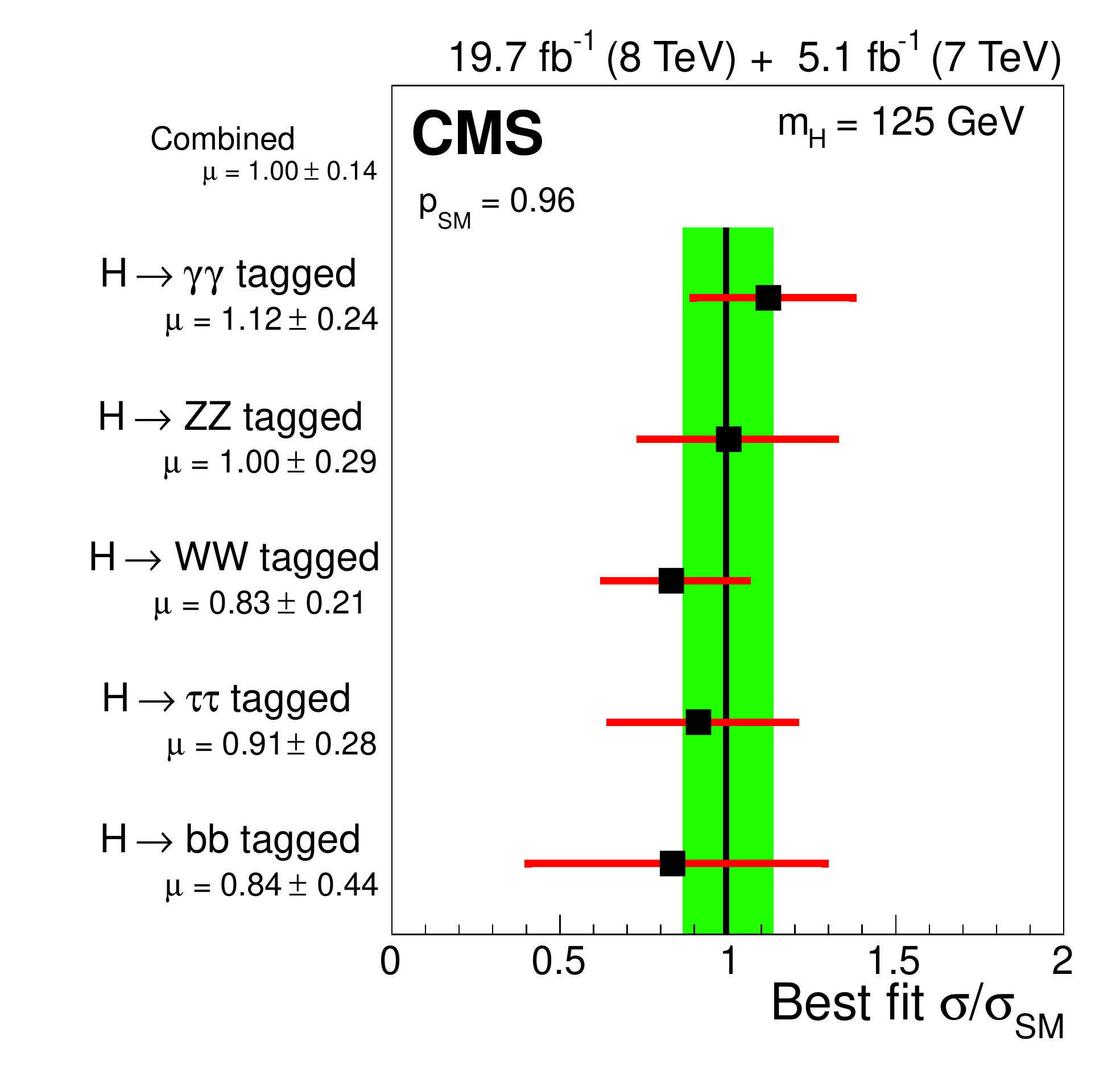}}
\caption{Left panel: ATLAS measured production strengths for a Higgs boson of mass $M_H =125.36\gev$, normalized to the standard-model expectations, for the prominent final states~\cite{Aad:2015gba}. The best-fit values are shown by the solid vertical lines. The total $\pm 1\sigma$ uncertainty is indicated by the shaded band.  Right panel: CMS values of the best-fit $\sigma/\sigma_{\mathrm{SM}}$ for the combination (solid vertical line) and for subcombinations by predominant decay mode~\cite{Khachatryan:2014jba}. The vertical band shows the overall $\sigma/\sigma_{\mathrm{SM}}$ uncertainty. The $\sigma/\sigma_{\mathrm{SM}}$ ratio denotes the production cross section times the relevant branching fractions, relative to the standard-model expectation. The horizontal bars indicate the $\pm 1 \sigma$ uncertainties in the best-fit $\sigma/\sigma_{\mathrm{SM}}$ values for the individual modes; including statistical and systematic uncertainties. \label{fig:strengths}}
\end{figure}

The $WW$ and $ZZ$ modes speak directly to the role of the new particle in electroweak symmetry breaking. \query{Since it carries no electric charge, the Higgs boson does not couple directly to photons. The $\gamma\gamma$ decay mode arises when $H$ fluctuates into a pair of charged particles, notably $W^+W^-$ or $t\bar{t}$, which do couple to photons. The decay rate is sensitive to the presence of new particles, beyond the expected $W$ and $t$, that may contribute to the one-loop Feynman diagram.} Characteristics of the production imply that the new particle couples to top quark pairs. Taken together with the evidence for $b\bar{b}$ and $\tau^+\tau^-$ decays, this constitutes evidence for Yukawa couplings that may be responsible for giving mass to fermions. The observed production rates for the prominent decay modes are collected in Figure~\ref{fig:strengths}.
Both ATLAS~\cite{ATLAS:2013qma} and CMS~\cite{Khachatryan:2014aep} have set 95\% Confidence Level upper limits on the $H \to \mu^+\mu^-$ branching fraction of approximately $2 \times 10^{-3}$, roughly $1/30 \times$ the $H \to \tau^+\tau^-$ level. Together with the nonobservation of the $e^+e^-$ decay, this confirms that the leptonic couplings of $H(125)$ are not flavor-universal, in agreement with the standard-model expectation that the Higgs-boson coupling to fermion pairs should be proportional to the fermion mass.\query{\footnote{\query{The electroweak theory attributes the masses of quarks and charged leptons to their interactions with the Higgs field.}}}

Once the existence of a new state is established, the next step is to characterize its quantum numbers. A particle that decays into two photons must be even under charge conjugation.\footnote{Charge conjugation invariance is treated in modern introductory textbooks, such as Refs.~\cite{TullyNutshell,DJGParticles}.} On the assumption that the two decay modes represent decays of the same new particle, observation of the $\gamma\gamma$ mode means, by the selection rule known as the Landau~\cite{Landau:1948kw}--Yang~\cite{Yang:1950rg} theorem, that the disintegrating particle cannot have spin one.\footnote{See Ref.~\cite{PhysRevA.91.033417} for generalized selection rules, with applications to atomic physics.} Extensive studies of the $H \to \gamma\gamma$, $H \to ZZ \to \ell^+\ell^-\ell^{\prime+}\ell^{\prime-}$, and $H \to WW \to \ell\nu \ell^\prime\nu^\prime$ channels decisively favor the spin-parity $J^P = 0^+$ assignment given by the electroweak theory, and rule against exotic $J^P = 0^-, 1^+, 1^-$ alternatives and a variety of $2^+$ hypotheses~\cite{Aad:2015mxa,Khachatryan:2014kca}. Some of the techniques used were developed long ago to determine the parity of $\pi^0$ in $e^+e^-e^+e^-$ ``Double-Dalitz'' decays.\footnote{The best modern determination, reported by the KTeV experiment at Fermilab, limits an admixture of scalar contributions in the decay amplitude to $3.3\%$~\cite{Abouzaid:2008cd}.}  Analyzing their excess $b\bar{b}$ events around $M_H$, the Tevatron experiments CDF and D0 strongly disfavor $J^P = 0^-$ and $2^+$ assignments~\cite{Aaltonen:2015mka}.  

The total width of the standard-model Higgs boson is calculated to be $\Gamma_H(M_H=125.09\gev) = 4.08\mev$~\cite{Heinemeyer:2013tqa}, which is far too small to be directly observed in the LHC environment. By comparing signal strengths \query{near the resonance peak and in the high-mass tail}, CMS~\cite{Khachatryan:2014iha} and ATLAS~\cite{Aad:2015xua} have inferred an upper bound on the Higgs-boson width of $\Gamma_H \lesssim 22\mev$ at 95\% confidence level. This intriguing methodology merits further development and scrutiny.

\subsection{What We Want to Learn \label{sub:sec:whatlearn}}
\begin{figure}
\centerline{\includegraphics[height=0.15\textheight]{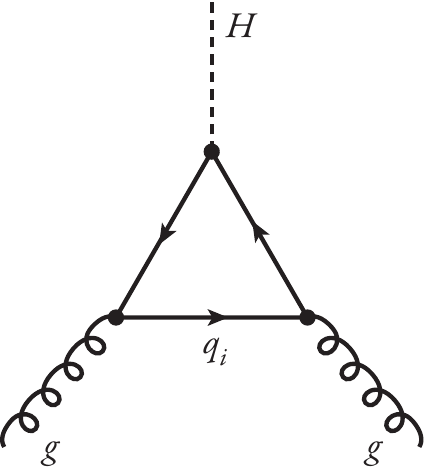}\qquad
\includegraphics[height=0.2\textheight]{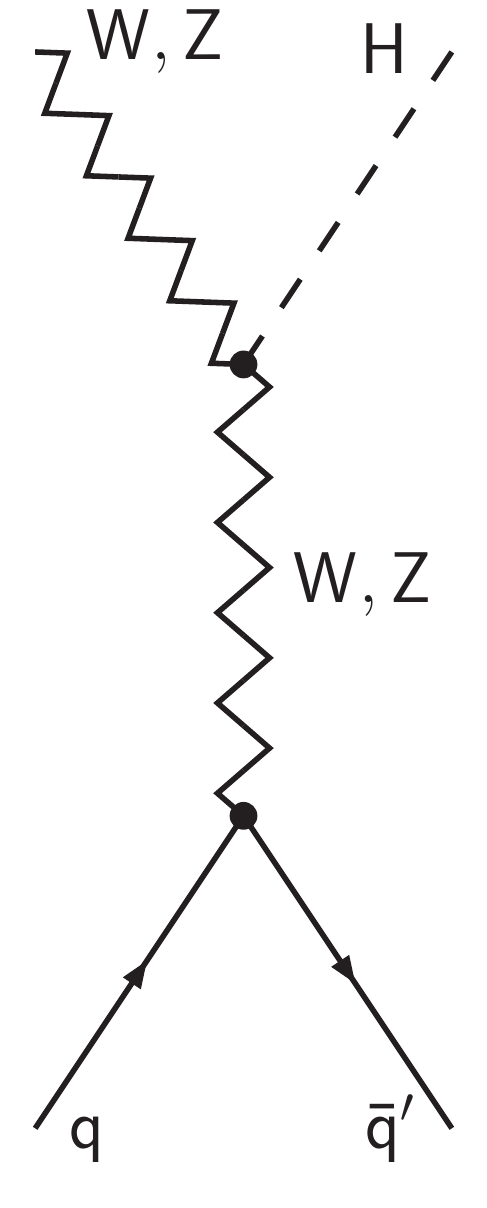}\qquad
\includegraphics[height=0.15\textheight]{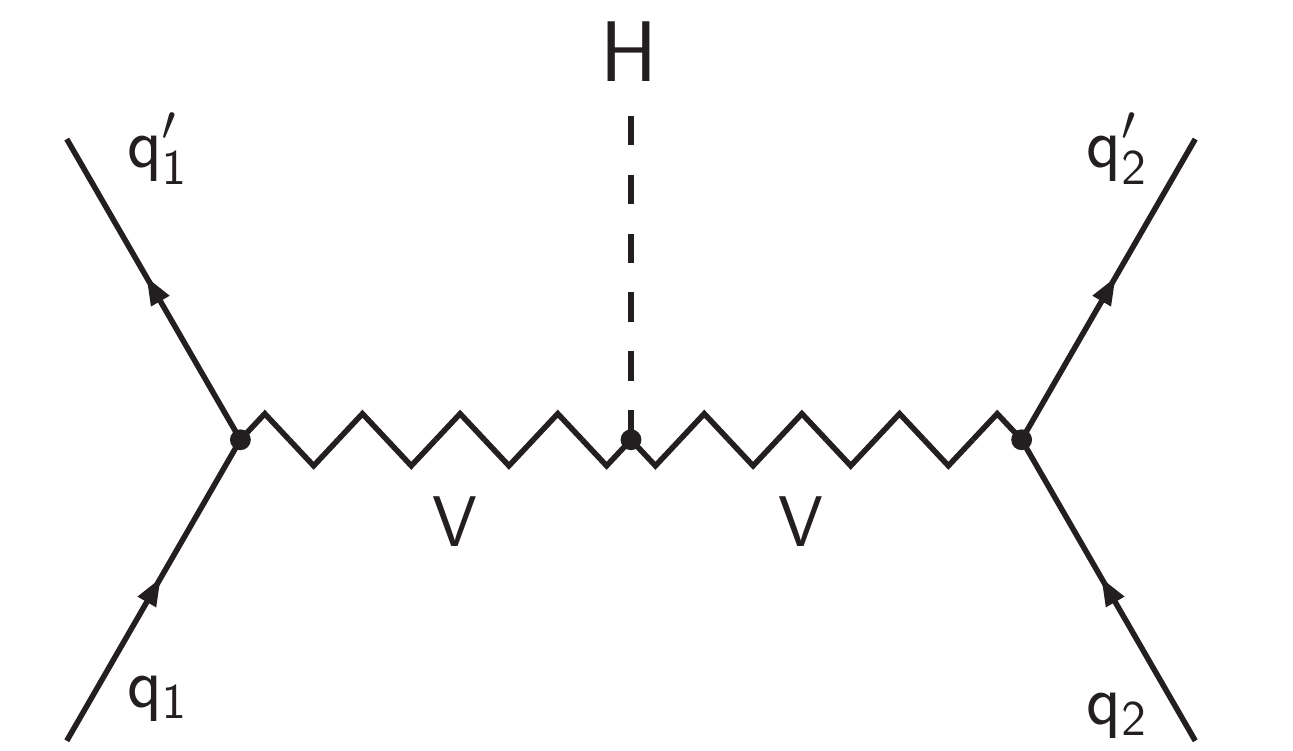}\qquad\includegraphics[height=0.15\textheight]{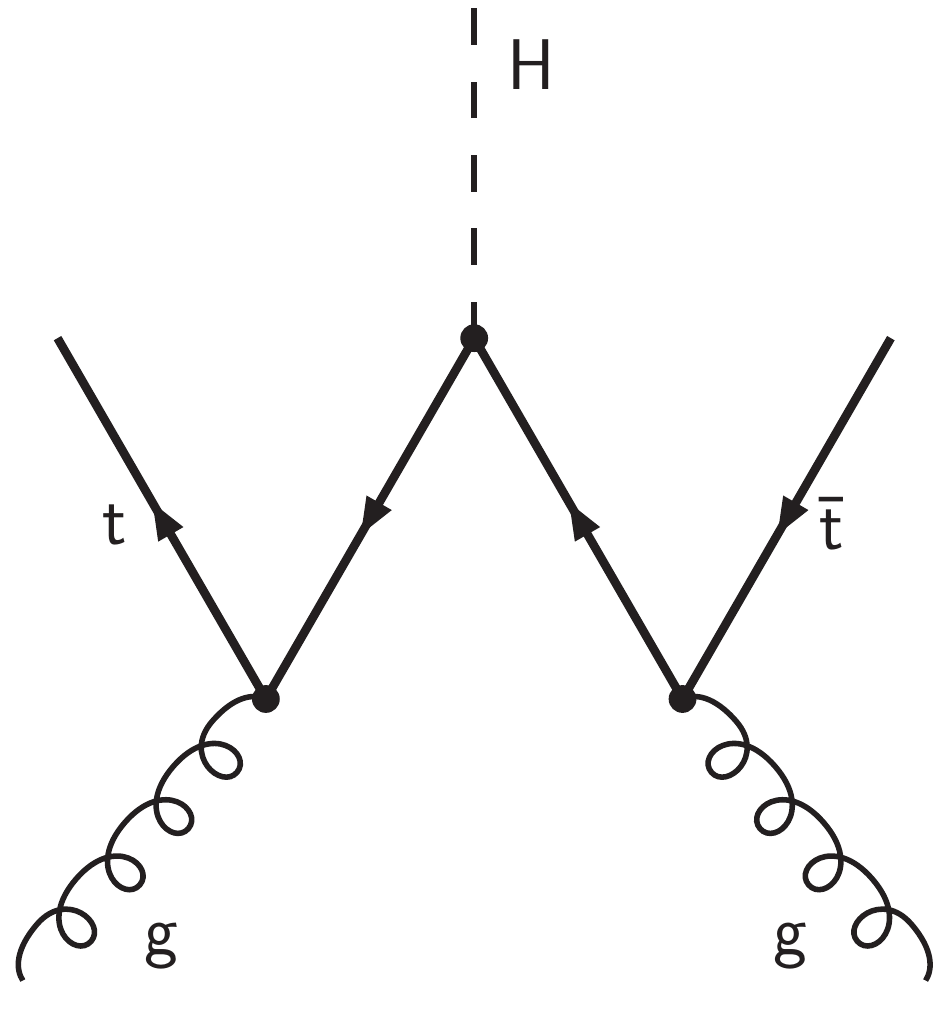}}
\caption{Principal Higgs-boson production mechanisms at the Large Hadron Collider. \label{fig:prodmech}}
\end{figure}

The best compact summary we can give about $H(125)$ is that evidence has been developing as it would for a standard-model Higgs boson---which does not mean that the verdict is settled. In Run~1 of the Large Hadron Collider, ATLAS and CMS observed Higgs-boson signals produced by the first three mechanisms shown in Figure~\ref{fig:prodmech}: gluon fusion through heavy-quark (top) loops, associated production with weak gauge bosons, and vector-boson fusion. 
Under plausible assumptions, the experiments have been able to characterize the couplings of $H(125)$ to gauge bosons and to fermions, including the $Ht\bar{t}$ coupling inferred from the gluon-fusion production rate~\cite{Khachatryan:2014jba,ATLAS:2015bea}. It is hoped that, in Run~2, observation of the $Ht\bar{t}$ production mechanism (rightmost diagram in Figure~\ref{fig:prodmech}) will allow a direct determination of the Higgs-boson coupling to the top quark.

\begin{figure}
\centerline{\includegraphics[height=0.3\textheight]{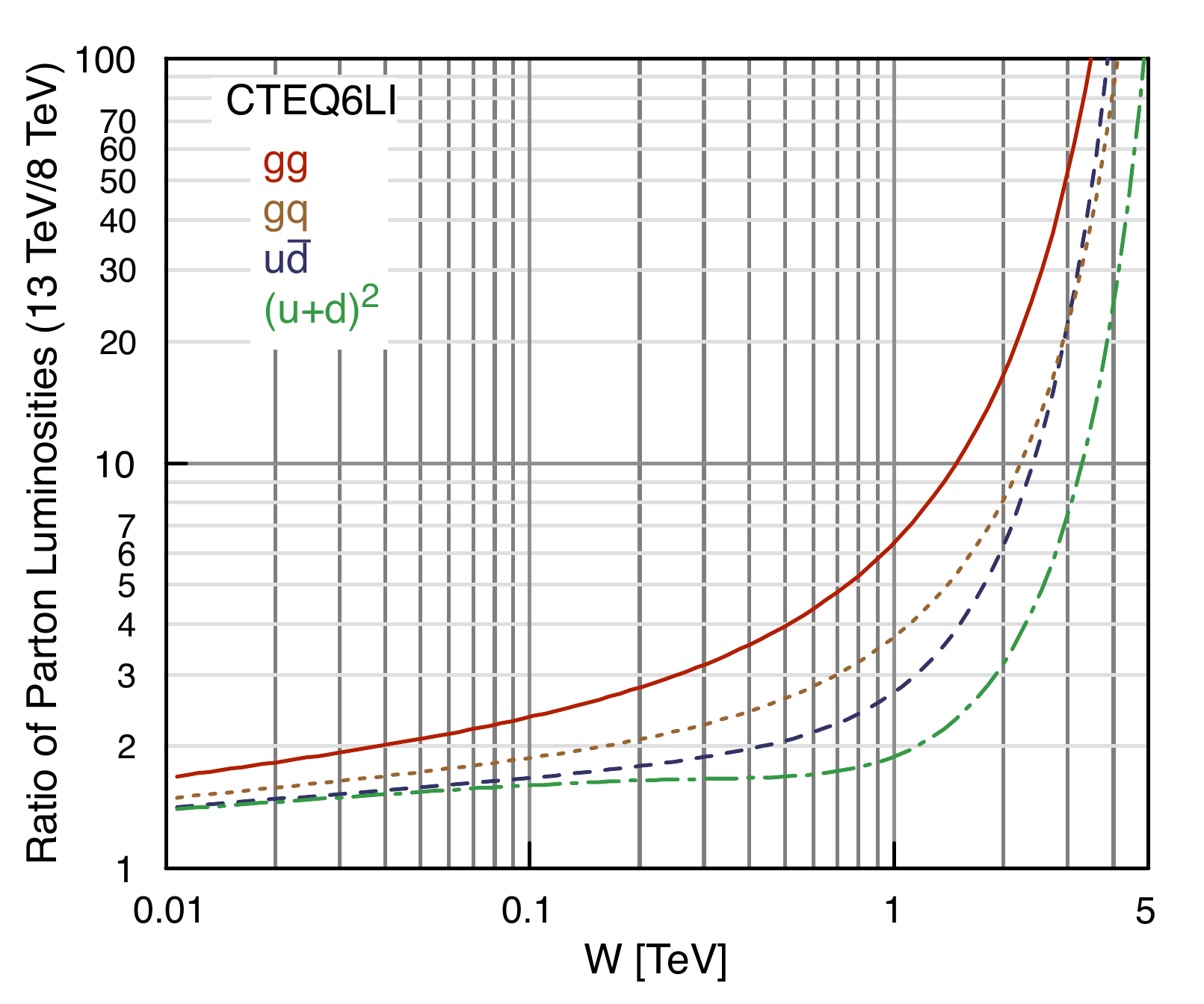}}
\caption{Ratios of parton luminosities at $\sqrt{s}=13\hbox{ and }8\tev$ as functions of the subenergy of the colliding partons. \label{fig:partonlums}}. 
\end{figure}

The run that is beginning in 2015 will be carried out at $\sqrt{s} = 13\tev$, and aims at an integrated luminosity of $300\fb^{-1}$, an order of magnitude larger than in Run~1. At this higher energy, the rate of parton-parton collisions\query{\footnote{\query{Parton is a generic term, introduced by Richard Feynman, for the constituents of a proton, which we now know to be quarks, antiquarks, gluons, and possibly other particles.}}} at a given effective mass, $\mathsf{W}$, is greater than at $\sqrt{s} = 8\tev$ by a factor that we can estimate from our knowledge of proton structure. I display such an estimate, based on the CTEQ6L1 parton distribution functions~\cite{Pumplin:2002vw},  in Figure~\ref{fig:partonlums} 
for four representative parton combinations: gluon--gluon, gluon--light quark, up--antidown, and light quark--light quark. We observe at once that the  rate of $gg$ collisions at $\mathsf{W} = 125\gev$ in $pp$ collisions at $\sqrt{s} = 13\tev$ (which governs the yield of Higgs bosons produced through gluon fusion)  is $2.5\times$ the rate at $\sqrt{s} = 8\tev$. Thus the first $10\fb^{-1}$ recorded in Run~2 will roughly double the world sample of Higgs bosons.

\begin{figure}
\centerline{\includegraphics[height=0.3\textheight]{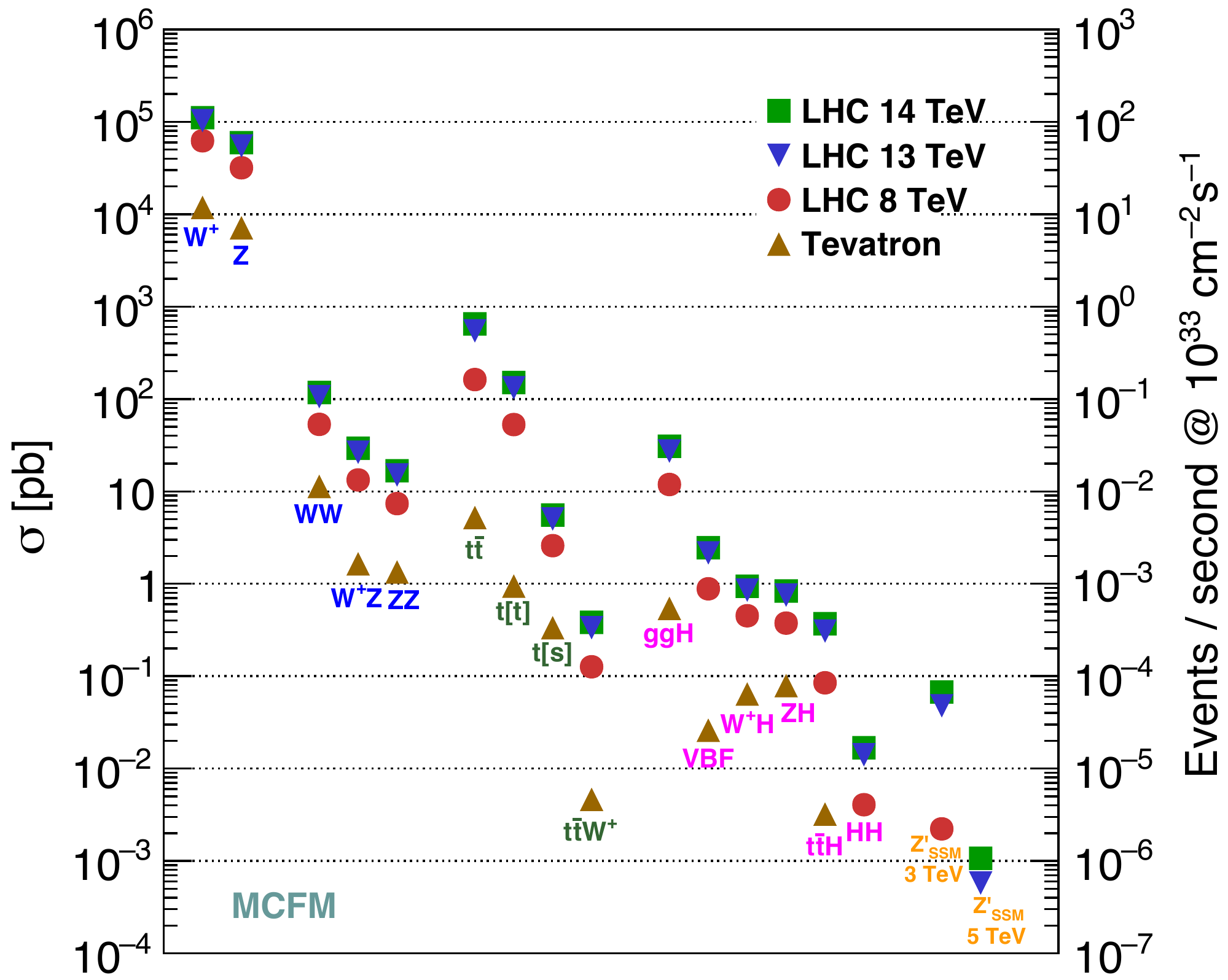}}
\caption{Calculated cross sections for reactions involving weak bosons and Higgs bosons in $\bar{p}p$ collisions at $\sqrt{s} = 1.96\tev$ and in $pp$ collisions at $\sqrt{s} = 8, 13, 14\tev$, calculated using \textsf{MCFM}~\cite{mcfm}. \label{fig:mcfmsigs}}
\end{figure}

A more direct representation of how rates for a variety of processes involving weak bosons and Higgs bosons evolve with energy appears in Figure~\ref{fig:mcfmsigs}. 
The entries for Higgs-boson production through gluon fusion ($ggH$), vector-boson fusion (VBF), and associated production of Higgs bosons with weak bosons ($W^+H, ZH$) make clear how challenging was the Higgs search at the Tevatron. The step from $\sqrt{s} = 8\tev\hbox{ to }13\tev$ is significant for Higgs-boson production, but reaching the design energy of $\sqrt{s} = 14\tev$ would bring only a slight advantage for those reactions. For the important $Ht\bar{t}$ final state, the cross sections at $\sqrt{s} = 8, 13, 14\tev$ are in the proportions $1:3.95:4.74$~\cite{Dittmaier:2011ti}.

Higgs-bearing events are rare: the total $pp$ cross section at the LHC is approximately $10^{11}\pb$, many orders of magnitude greater than the Higgs-boson production rates, even before the expected decay probabilities, shown in Table~\ref{tab:HBRs}, are taken into account.\footnote{The inelastic and elastic cross sections are approximately $3/4$ and $1/4$ of the total, respectively.} 
\begin{table}
\tbl{Branching Fractions for a Standard-Model Higgs Boson with $M_H =125.1\gev$~\cite{Heinemeyer:2013tqa}. \label{tab:HBRs}}
{\begin{tabular}{cc}
\toprule
Higgs-boson decay mode & Branching fraction \\
\colrule
$b\bar{b}$ & 0.575 \\
$W^+W^-$ & 0.216 \\
$gg$ & 0.0856 \\
$\tau^+\tau^-$ & 0.0630 \\
$c\bar{c}$ & 0.0290 \\
$Z^0Z^0$ & 0.0267 \\
$\gamma\gamma$ & $2.28 \times 10^{-3}$ \\
$\gamma Z^0$ & $1.55 \times 10^{-3}$ \\
$s\bar{s}$ & $2.46 \times 10^{-4}$ \\
$\mu^+\mu^-$ & $2.19 \times 10^{-4}$ \\
$e^+e^-$ & $5.12 \times 10^{-9}$\\
\botrule
\end{tabular}}
\end{table}
Thus high proton-proton luminosity is required of the collider and swift and well-conceived event selection is required of the detectors.

A prime goal of LHC Run~2 is to complete the $H(125)$ dossier, and to determine how closely the Higgs boson conforms to the predictions of the \ewgg\ electroweak theory. One important element will be to pin down the branching fractions---even better, the absolute decay rates---or equivalently the couplings to gauge bosons and fermions with improved precision.\footnote{If it can achieve adequate luminosity, an $e^+e^- (\to HZ)$ ``Higgs factory'' would be, for some purposes, superior to the LHC environment. The best-developed science case is that prepared for the International Linear Collider project~\cite{Moortgat-Pick:2015yla,Fujii:2015jha}.} A central issue is whether the $HZZ$ and $HWW$ couplings indicate that $H(125)$ is the sole agent of electroweak symmetry breaking. We want to consolidate the indications that $H(125)$ plays a part in giving mass to fermions, specifically $t, b, \tau$, and test whether interactions with the $H$ field fully account for those masses. It is then essential to discover whether this mechanism also applies to the lighter fermions, or whether they might derive their masses by some sort of quantum  corrections that arise from the heaviest fermion masses. 

In the foreseeable future, only the decays into ``second-generation'' fermions seem within reach, and of those there is a clear prospect only for observing $H \to \mu^+\mu^-$. A credible strategy for observing Higgs decay into charmed-quark pairs would be a welcome advance. If we can show that spontaneous electroweak symmetry breaking is indeed the origin of fermion masses, that will give us a new understanding of how compact atoms, valence bonding, and composite structures such as solids and liquids can exist~\cite{Quigg:2009xr}. To complete this demonstration, we will need to determine the Higgs-boson coupling to electrons, which is (see the final entry in Table~\ref{tab:HBRs}) extraordinarily demanding. If the Higgs-fermion Yukawa couplings do indeed account for the fermion masses, the origin of the Yukawa couplings is still mysterious.

Along with the $\gamma\gamma$ discovery mode and the $Hgg$ effective coupling that governs the gluon-fusion production rate, the $H \to \gamma Z$ channel is induced by a one-loop Feynman diagram, so is sensitive to hitherto unknown particles that couple to the Higgs boson. At one more remove from established particle-physics phenomenology, the Higgs boson might couple to new forms of matter that do not otherwise interact with standard-model particles, perhaps through new interactions. The search for ``invisible'' decays, as well as other exotic decays, of $H(125)$ presents many opportunities for discovery~\cite{Curtin:2013fra}. According to the standard model, Higgs-boson decays into fermions should be flavor-neutral. For example, $H \to \tau^+\tau^-$ and $H \to \mu^+\mu^-$ are allowed, but $H \to \tau^\pm\mu^\mp$ is not. Searches for such forbidden decays, already initiated in Run~1, may turn up surprises.

In the standard electroweak theory, electroweak symmetry breaking is the work of a single complex doublet of scalar fields. It is easy to imagine generalizations, in particular to suppose that $H(125)$ might have partners that contribute in some measure to electroweak symmetry breaking. ATLAS and CMS have conducted broad searches for heavier partners of $H(125)$, and these will continue with higher sensitivity. These direct searches complement the determination of $HZZ$ and $HWW$ couplings. In specific models, including supersymmetry and other two-Higgs-doublet models, five Higgs bosons are implied: two neutral scalar states, one of which may closely resemble the standard-model Higgs boson, a neutral pseudoscalar, and a pair of charged scalars. Direct searches are ongoing. Looking for evidence of a pseudoscalar admixture in the dominantly $J^P = 0^+$ $H(125)$ also probes for additional degrees of freedom or for new strong dynamics in the Higgs sector.

Each of the production mechanisms depicted in Figure~\ref{fig:prodmech} can provide specific information about $H(125)$ and the particles to which it couples. From gluon--gluon fusion, we derive information about the $Ht\bar{t}$ coupling and possibly evidence for new heavy particles in the triangle. Indeed, the rough agreement between observation and standard-model expectation in Run~1---along with null results from many direct searches---severely constrains a fourth generation of quarks, for example.   The $H(W,Z)$ associated-production and vector-boson-fusion reactions test the $HWW$ and $HZZ$ couplings, whereas the $Ht\bar{t}$ reaction probes the Higgs-boson coupling to the top quark.

Two important questions are almost certainly out of reach for the LHC, even in the eventual High-Luminosity run that aims to accumulate $3\,000\pb^{-1}$ at $\sqrt{s} \gtrsim 13\tev$. If the $HWW$ couplings deviate from the standard-model value, then the $W_0W_0$ cross section, where the subscript $0$ denotes the longitudinal component, should increase rapidly with the $WW$ c.m.\ energy. Given the constraints already in hand and the modest $W_0W_0$ luminosity expected at LHC energies, progress from this quarter seems unlikely in the near future~\cite{Campbell:2015vwa}. As a consistency check, it would be interesting to measure the $HHH$ self-coupling, which is in principle accessible as a component of the reaction $pp \to HH + \hbox{anything}$. A measurement with meaningful uncertainty is likely to require a future high-luminosity collider---either a ``100-TeV'' $pp$ machine or a multi-TeV $e^+e^-$ machine. The two issues discussed in this paragraph set worthy targets for energy-frontier colliders of the next generation.

The value of the Higgs-boson mass has implications for specific models that posit physics beyond the standard model. Moreover, it speaks to the range of applicability of the electroweak theory. Within the electroweak theory as we have formulated and tested it, it appears that we may live in a false \query{(metastable)} vacuum in which both $M_H$ and $m_t$ have near-critical values~\cite{Buttazzo:2013uya}. Are we living on borrowed time, or is our vacuum stabilized by new physics?

\section{Beyond the Higgs Boson \label{sec:notHiggs}}
The first task for experiments in a new regime of energy and sensitivity is to rediscover existing knowledge, to validate the instruments and begin to make increasingly incisive measurements. Early on, it will be possible to extend the search for quark and lepton compositeness, \emph{checking the foundations of the standard model.} A contact interaction, which should be the first signature of constituents that have finite size and internal structure, can alter the cross section for dijet or dilepton production, and cause production angular distributions to deviate from standard-model expectations~\cite{Eichten:1983hw}. The discovery of free quarks would be revolutionary, precisely because it is so unlikely in the context of our understanding of Quantum Chromodynamics, the theory of strong interactions~\cite{Kronfeld:2010bx}.

\begin{figure}
\centerline{\includegraphics[height=0.4\textheight]{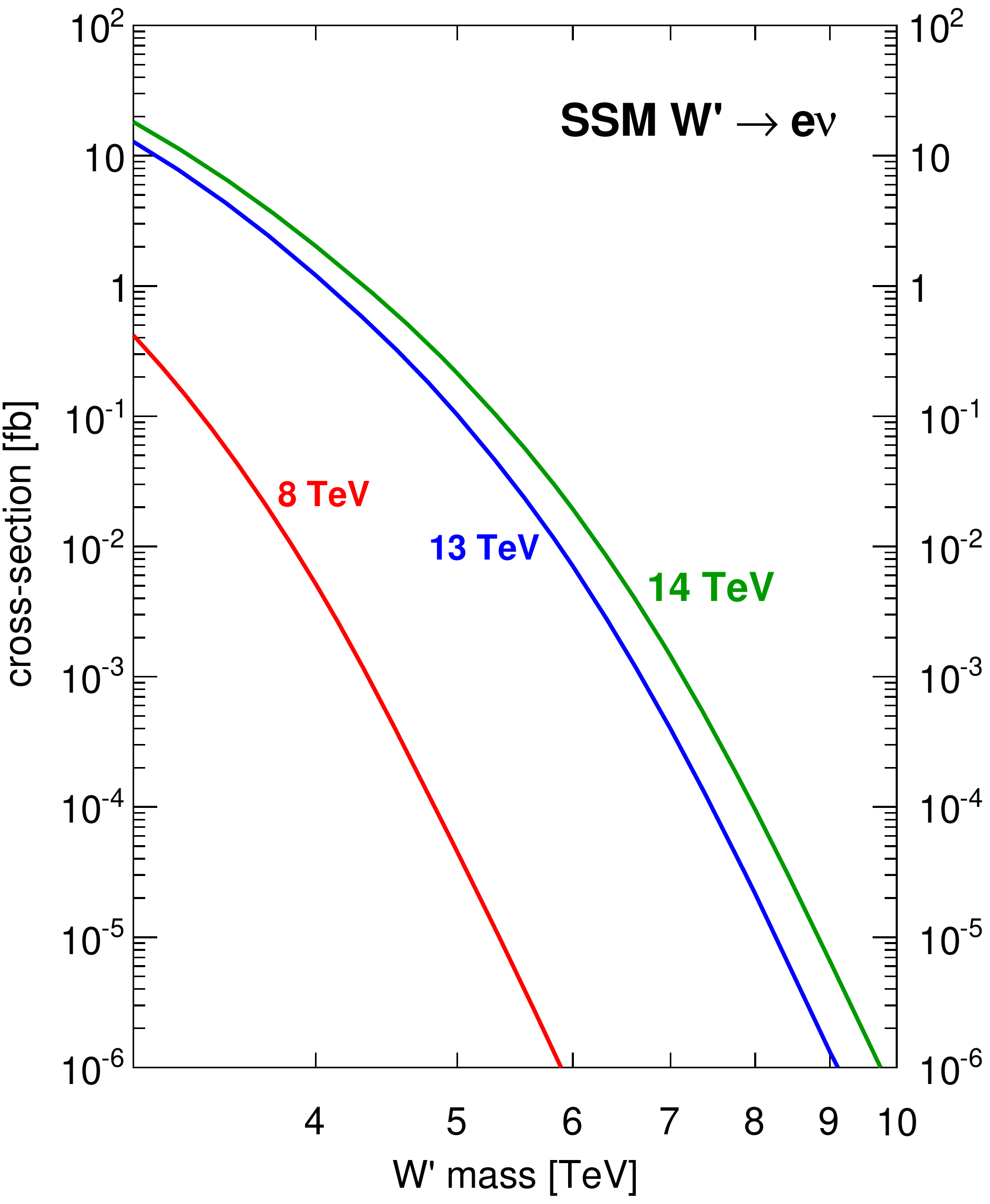}\quad\includegraphics[height=0.4\textheight]{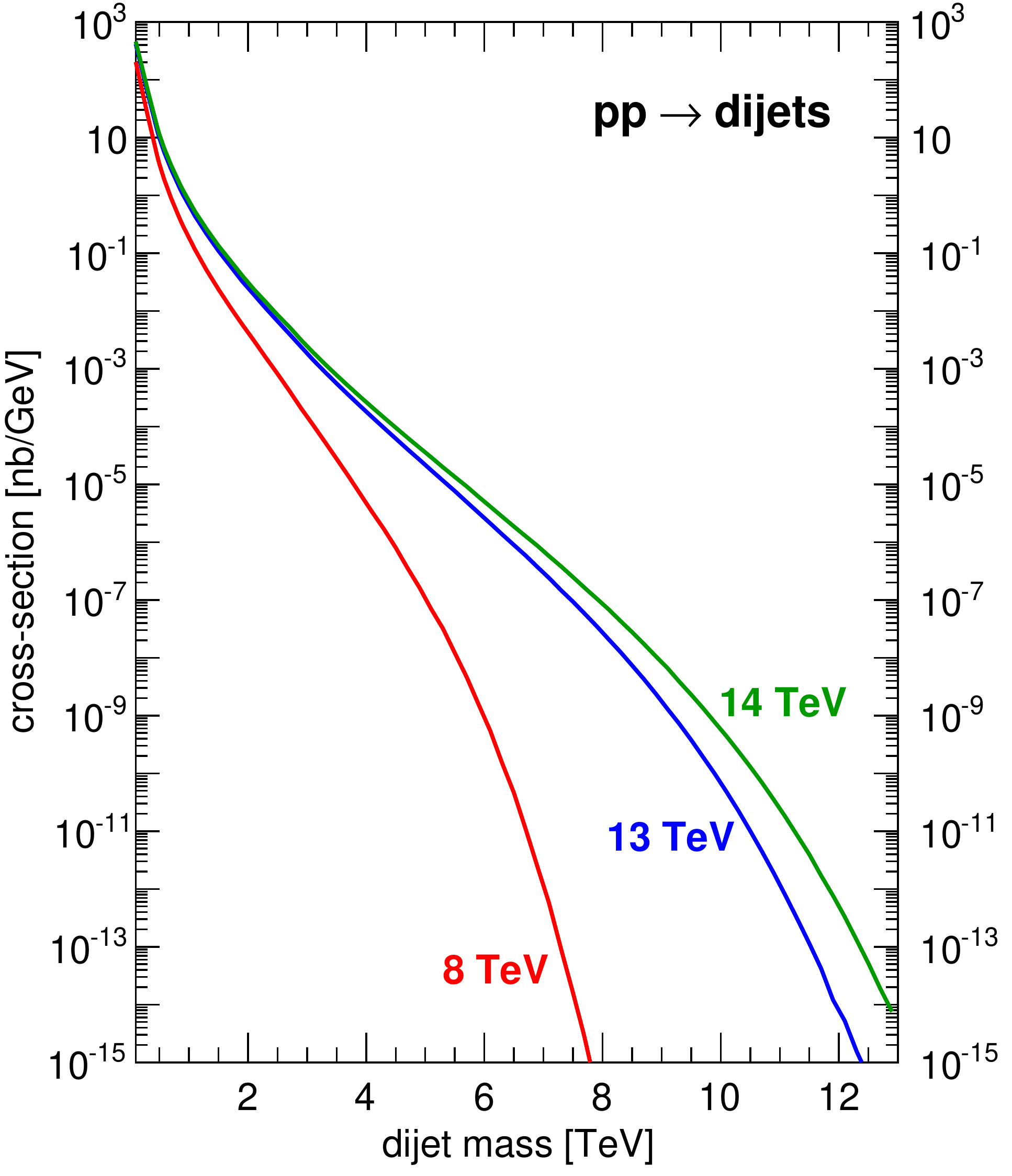}}
\caption{Left panel: Cross sections for the production of $W^\prime \to e\nu$ with standard-model couplings in $pp$ collisions at $\sqrt{s} = 8, 13, 14\tev$. Right panel: Cross sections for dijet production in $pp$ collisions at $\sqrt{s} = 8, 13, 14\tev$.  \label{fig:combined}}
\end{figure}

I classify as ``not unexpected'' new weak gauge bosons, because it is easy to imagine generalizations---either \emph{ad hoc} or following from a unified theory of the fundamental interactions---of the \ewgg\ electroweak gauge group. An artificial, but nevertheless useful, construct is a sequential-standard-model gauge boson, one with the same couplings as the corresponding \ewgg\ particle. Limits from current searches are given in the 2014 \emph{Review of Particle Physics}~\cite{Agashe:2014kda}: $M_{W^\prime} > 2.90\tev$ and $M_{Z^\prime} > 2.59\tev$, both at 95\% CL. Predicted cross sections for $W^{\prime} \to e\nu$ production in $pp$ collisions at $8, 13,\hbox{ and }14\tev$, calculated with \textsf{MCFM}~\cite{mcfm}, are shown in the left panel of Figure~\ref{fig:combined}. 
At $M_{W^\prime} = 3\tev$, the cross section rises by nearly a factor of 30 when the collision energy increases to $\sqrt{s} = 13\tev$ (\cf the $u\bar{d}$ luminosity ratio in Figure~\ref{fig:partonlums}). This means that the Run~1 sensitivity will be equalled with the accumulation of $< 1 \fb^{-1}$ at $\sqrt{s}=13\tev$. The signal of an isolated, high-transverse-momentum lepton with missing transverse momentum opposite is essentially background-free. Thus we may estimate that $30\fb^{-1}$ at $13\tev$ gives sensitivity to $M_W$ somewhat larger than $4\tev$, and that $300\fb^{-1}$ extends the reach by another TeV or more. For an exposure of $300\fb^{-1}$ at $14\tev$, explicit studies by ATLAS~\cite{ATLAS:2013hta} suggest that the 95\% CL exclusion reach for a sequential $Z^\prime$ is $6.5\tev$;  CMS studies~\cite{CMS:2013xfa} set the 5-$\sigma$ discovery limit at $5.1\tev$, at an integrated luminosity of $30\fb^{-1}$.

A broader class of new force particles that I would characterize as ``imagined,'' may be sought by direct production or through their effect on familiar processes. Each of these entails its own motivation and search strategy, which we need not discuss here. Examples include axigluons, massive mediators of a strong force that would arise when a larger (chiral) color symmetry is broken down to Quantum Chromodynamics, $\mathrm{SU(3)_L \otimes SU(3)_R \to SU(3)_c}$~\cite{Frampton:1987dn}; colorons \query{(massive color-octet gauge bosons)}, or other manifestations of new strong dynamics~\cite{Dynamics}; leptoquarks \query{(particles coupling leptons to quarks)} that arise in unified theories or in dynamical-symmetry breaking pictures such as technicolor\query{, in which electroweak symmetry is hidden by the strong interactions of a new gauge force}; or Ka\l uza--Klein recurrences \query{(heavier copies of standard-model particles)} or other manifestations of extra spatial dimensions~\cite{ED}. 

Motivations and strategies for resonance searches in dijet final states are presented in Ref.~\cite{Dobrescu:2013cmh}. Quark-quark scattering dominates the production of the highest invariant-mass dijets at the LHC. We see from the parton luminosity ratios in Figure~\ref{fig:partonlums} that the advantage of $13\tev$ over $8\tev$ becomes pronounced for dijet masses $\mathsf{W} \gtrsim 2\tev$. The highest dijet mass recorded at $\sqrt{s} = 8\tev$ was $\mathsf{W} = 5.15\tev$, in the CMS experiment. The right panel of Figure~\ref{fig:combined} shows that the cross section calculated with \textsf{MCFM}~\cite{mcfm} is approximately $400\times$ larger at $\sqrt{s} = 13\tev$, which suggests that Run~2 should begin to break new ground at an integrated luminosity of less than $0.1\fb^{-1}$. With $30\fb^{-1}$ of data, this rough measure of single-event sensitivity grows to $\sim 7.7\tev$, and with $300\fb^{-1}$, to perhaps $8.6\tev$.  These numbers would be approximately 10\% larger at $\sqrt{s} = 14\tev$. Among conceivable dijet resonances, it is well to remember diquark resonances, even if the best motivation I can supply is ``Why not?''

It is worth underlining that establishing a new force of nature or finding that our ``fundamental'' constituents are composite would be a major discovery.

In the next category, ``long awaited,'' I place the creation of dark-matter candidates in the laboratory. If dark matter is composed of thermal-relic particles that interact weakly with ordinary matter, then (assuming that a limited number of species make up the dark matter), the observed dark-matter density in the current universe, about 26\% of the critical density~\cite{Adam:2015rua}, points to dark-matter particles with masses ranging from a few hundred GeV to $1\tev$ or so. This is prime territory for searches during LHC Run~2~\cite{DMbench}. It is worth emphasizing that the discovery of candidates at the LHC cannot establish that the new particles are stable on cosmological time scales. That is the business of direct and indirect dark-matter searches. The strength of collider experiments will lie in detailing the properties of the dark-matter candidates. It would be a wonderful achievement to characterize a major, and as yet unidentified, component of  matter in the universe at large.
 
 Even after its apparent completion by the observation of a light Higgs boson, the electroweak theory raises puzzles. An outstanding question is why the electroweak scale is so much smaller than other plausible physical scales, such as the unification scale and the Planck scale. In other words,  what stabilizes the Higgs-boson mass below $1\tev$, when it seems natural for the weak scale to be comparable to other physical scales---a unification scale around $10^{16}\gev$ or the Planck scale near $10^{19}\gev$? Among so-called ``natural'' explanations~\cite{Dine:2015xga}, TeV-scale supersymmetry is the stuff of many theorists' dreams~\cite{Ellis:2015daa}. Extensive searches for superpartners of the standard-model particles were carried out in Run~1. It is plausible that in Run~2 limits (or discoveries!) can be pushed to $1\tev$ for the top squark, $\tilde{t}$, the spin-$0$ partner of the top quark, and to $1.5\tev$ for the gluino, $\tilde{g}$, the spin-$\frac{1}{2}$ superpartner of the gluon. Searches for the lightest supersymmetric particle (generally presumed stable), like searches for dark-matter particles, typically key on missing transverse energy. Backgrounds are a concern, and the limits set by specific searches always have evasions.
 
 In principle, a global search for new colored degrees of freedom, such as squarks and gluinos, can be conducted by measuring the scale dependence of $\alphas$ the strong-interaction analogue of the fine structure constant. In quantum field theory, coupling strengths depend on the momentum scale on which they are measured. The variation of $\alpha$ itself is familiar in Quantum Electrodynamics as a consequence of vacuum polarization. Virtual electron--positron pairs screen a test charge---as virtual particles with charge opposite in sign to the test charge are attracted and those with the same-sign charge are repelled. In consequence, the effective charge increases at short distances, or high momentum scales.  In pure QED---the theory of electrons and photons---the variation (at leading order) is 
 ${1}/{\alpha(Q)} = {1}/{\alpha(\mu)} - ({1}/{6\pi})\ln\left({Q}/{\mu}\right)$, where $Q$ is the momentum scale of interest and $\mu$ is a reference scale. 
 
 In QCD, the evolution of \alphas\ is influenced both by an analogous screening and by antiscreening that arises from the fact that gluons carry color charge. The result of the competition is a net antiscreening that can be characterized (at leading order) for momenta approaching $1\tev$ as $1/\alphas(Q) = 1/\alphas(\mu) + (7/2\pi)\ln\left({Q}/{\mu}\right)$. This is the celebrated property of \emph{asymptotic freedom,} the tendency of \alphas\ to become small at high momentum scales or short distances, which is the basis for perturbative calculations of hard-scattering processes\query{---large--momentum-transfer collisions of partons}. If a full set of virtual squarks and gluinos were to come into play  on a certain mass scale---the traditional expectation was $1\tev$---then above that scale the slope of $1/\alphas$ would change from $7/2\pi$ to $3/2\pi$.\footnote{See the discussion surrounding Table~2 of Ref.~\cite{CQGT2}.} Reliably determining \alphas\ at high scales will not be easy---it will require considerable experimental and theoretical effort---but either observing or ruling out such a change in slope can teach us something important about the prospects for TeV-scale supersymmetry and coupling-constant unification.

\section{Concluding Remarks \label{sec:conc}}
To make the most of the physics opportunities before us, I would counsel a three-fold approach to experimentation: Explore, Search, Measure! With the increase of energy from $\sqrt{s} = 8\hbox{ to }13\tev$, and with the prospect of greatly increased integrated luminosity, we are entering unexplored terrain. I am confident that it will prove very rewarding to spend some time simply \emph{exploring} the new landscapes, without strong preconceptions, to get the lay of the land, perhaps to encounter interesting surprises. Directed \emph{search} campaigns, such as the extensive Run-1 searches for supersymmetry, will be extended, but also broadened in scope over time. Incisive null results can help us to discard, refine, or modify theoretical speculations, or to change the way we frame key issues---perhaps the hierarchy problem\query{, the puzzle of why the electroweak scale (characterized by the $W$- and Higgs-boson masses) is so much smaller than either the Planck scale or the energy at which the strong, weak, and electromagnetic interactions might be unified}. As we have seen in our discussions of $W^\prime$ searches and dijets, the huge increase in sensitivity in the regime explored in Run~1 means that any tantalizing hint of a signal will either blossom into a discovery or be exposed as a statistical fluctuation in short order. Finally, we can learn just how comprehensive is our idealized---and highly successful---conception of particles and forces by making precise \emph{measurements} and probing for weak spots, or finding more sweeping accord between theory and experiment.

\section*{Acknowledgment}\label{ACK}
I thank John Campbell for informative discussions and for providing the plots in Figure~\ref{fig:combined}.
Fermilab is operated by Fermi Research Alliance, LLC, under Contract No. DE-AC02-07CH11359 with the United States Department of Energy. I am grateful to John Iliopoulos and the \textit{Fondation Meyer pour le d\'eveloppement culturel et artistique} for generous support in Paris.
%
\noindent  
 \bibliographystyle{h-physrev5}
\bibliography{AfterHiggs}

\label{lastpage}

\end{document}